# Quantum like representation of a Spiral Phase Plate

**Fabio Antonio Bovino.**

*Quantum Optics Lab, Selex-SI, via Puccini 2, Genova, 16154, ITALY.*
*[\*]fbovino@selex-si.com*

**Abstract:** We introduce a quantum like representation of a Spiral Phase Plate, acting on an electromagnetic field, as a two mode phase operator. The representation is based on the Newton binomial expansion and on properties of rational power of lowering and raising operators of quantum field. The correctness of this representation is proved by obtaining the same results of the Paul's operator in the single mode limit and comparing the results of two particular problems solved both in the classical and quantum picture: the action of a Spiral Phase Plate on a Gaussian Beam (corresponding to the vacuum state of the two-dimensional harmonic oscillator) and on a off-axis Gaussian Beam (corresponding to the displaced vacuum state in quantum picture).

## 1. Introduction

Optical vortices (OV) in light beams are tightly bound to phase dislocations (or singularities): due to the continuous spatial nature of a field, the presence of these defects implies the vanishing of the field's amplitude in the singularity. In some types of dislocations the phase circulates around the singularity and creates a vortex [1]. Nye and Berry [2] used the term "phase dislocation" to define the locus of the zero amplitude of a field. Similarly to crystallography, phase dislocations can be classified in edge, screw and mixed screw-edge [3]. The interest in OV increased because of the fact that fields in which they are included show an helical wave-front structure - developing around the screw dislocation line – entailing the presence of an orbital angular momentum (OAM). This is an important feature, which can be exploited in several applications, from optical tweezers to the generation of N–dimensional quantum states (q-nits) for quantum information applications.

A screw wave dislocation can be defined by means of the integer topological charge Q, which represents the phase winding number, and can be found through a circulation integral around the dislocation line:

$$Q = \frac{1}{2\pi} \oint df \qquad (1)$$

where *f* is the phase of the field.

The most used method of generating OAM in an optical beam is by imprinting one or more vortices on its transverse field distribution, in practice by passing a fundamental Gaussian beam through a device that modifies only the phase, such as a Spiral Phase Plate (SPP). When a Gaussian beam is diffracted off such an SPP, the resulting mode can be viewed as a superposition of Laguerre–Gaussian (LG) modes.

It is known that a Laguerre-Gaussian beam of paraxial light has a well-defined orbital angular momentum [4–7]. In Ref. [6] Nienhuis and Allen employed operator algebra to describe the Laguerre-Gaussian beam, and noticed that Laguerre-Gaussian modes are laser mode analog of the angular momentum eigenstates of the isotropic 2-d harmonic oscillator. In Ref. [10] Simon and Agarwal presented a phase-space description (the Wigner function) of the LG mode by exploiting the underlying phase-space symmetry. Li-Yun Hu and Hong-yi Fan have shown that LG mode is just the wave function of the common eigenvector of the orbital angular momentum operator and the total photon number operator of 2-d oscillator in the entangled state representation [11, 12], which is based on the first formulation of Einstein-Podolsky-Rosen quantum entanglement (EPR paradox).

In the Quantum Mechanics picture it is useful to search a quantum operational representation of a Spiral Phase Plate because, according to Dirac, there should exists a formal correspondence between quantum optics operators and classical optics transformations. Aiello et al [13] proposed a quantum operator representing the action of the SSP on a Gaussian Beam: it is the analog of the quantum phase operator.

The issue of defining a quantum phase operator for the electromagnetic field is a great challenge in quantum mechanics and quantum optics. Quite few proposals were made to define quantum phase operators consistent with quantum mechanics and coinciding with an experimental measurement. The most frequently discussed are those proposed by Susskind and Glogower [14], Pegg and Barnett [15], Paul [16], Noh, Fougires and Mandel [17], and Shapiro and Wagner (SW) [18]. The last ones are based on the simultaneous detection of quadrature components using a heterodyne detection method. The output of the heterodyne detection is associated with the measurement of the complex amplitude,

$$Y = \hat{a}_1 + \hat{a}_2^\dagger, \tag{2}$$

where *1* and *2* are signal and idler modes, respectively, with $\left[\hat{a}_1, \hat{a}_1^\dagger\right] = \left[\hat{a}_2, \hat{a}_2^\dagger\right] = 1$. Since $\hat{Y}^\dagger$ commutes with $\hat{Y}$, $\left[\hat{Y}, \hat{Y}^\dagger\right] = 0$. This point is essential in simultaneous measurements of the two quadrature components. Hradil [19, 20] then, pointed out that the detection of the same quantity, as given in Eq. (2), can be achieved with a double homodyne detection scheme. By extending the idea of the Susskind-Glogower's phase operator, introduced as a polar decomposition of the annihilation operator [14], he defined a phase operator (or phase-like variable) as

$$\hat{e}^{i\varphi} = \sqrt{\frac{\hat{Y}}{\hat{Y}^\dagger}} = \sqrt{\frac{\left(\hat{a}_1 + \hat{a}_2^\dagger\right)}{\left(\hat{a}_1^\dagger + \hat{a}_2\right)}}. \tag{3}$$

Fan Hong-Yi and Min Xiao, starting from the 2-d oscillator in the entangled state representation [21, 25], found an expression in which the two-mode phase operator $\hat{e}^{i\varphi}$ manifestly exhibits its phase behavior.

In this paper we express the quantum phase operator related to a SPP in a novel way, which allows its application directly to Fock States without using P representation. The paper is organized as follows: we summarize the properties of a Spiral Phase Plate and quantum algebra of paraxial beams. Then, we introduce the quantum representation of a Spiral Phase Plate and the connections with phase operator. Finally, we solve two diffraction problems by using classical and quantum approach to demonstrate the exactness of the method.

## 2. The Spiral Phase Plate

A Spiral Phase Plate (SPP) is a transparent plate of refractive index *n*, whose thickness is proportional to the azimuthal angle $\varphi$

$$h = h_s \frac{\varphi}{2\pi} + h_0, \tag{4}$$

where $h_s$ is the step height and $h_0$ the base height of the device (see Fig. 1). When such a plate is inserted in the waist of a Gaussian beam, where the phase distribution is plane, it imprints a vortex charge

$$q = \frac{h_s(n - n_0)}{\lambda}. \tag{5}$$

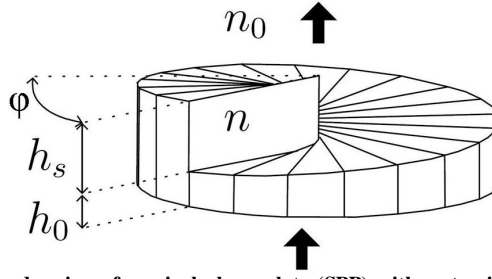

**Fig. 1: Schematic drawing of a spiral phase plate (SPP) with a step index $q=h_s(n-n_0)/\lambda$, where $h_s$ is the step height, n and $n_0$ are the refractive indices of the SPP and the surrounding medium, respectively, and λ is the wavelength of the incident light.**

The output beam thus carries orbital angular momentum per photon equal to $q\hbar$. The vortex charge that such a device imprints on an optical beam can be tuned by modifying either the optical step height $h_s(n-n_0)$, or the optical wavelength $\lambda$.

When a Gaussian beam is diffracted off such an SPP, the resulting mode can be viewed as a superposition of LG modes. The classical solution of the problem is well known. For a Gaussian Beam impinging on a SPP the output field is given by

$$V(r,\varphi,z) = \frac{2}{w(z)}\sqrt{\frac{1}{2\pi}}e^{[ikz-\Phi(z)]}e^{\left[-\frac{r^2}{w^2(z)}+\frac{ikr^2}{2R(z)}\right]}e^{iq\varphi}, \tag{6}$$

which, for $z=0$ can be expanded in the form

$$V(r,\varphi,0) = \sum_{p=0}^{\infty}\sum_{l=-\infty}^{\infty} C_{pl} u_{pl}^{LG}(r,\varphi,0), \tag{7}$$

where

$$u_{pl}^{LG}(r,\varphi,0) = \sqrt{\frac{2}{\pi w_0^2}}\sqrt{\frac{p!}{(p+|l|)!}}e^{-\frac{r^2}{w_0^2}}\left(\frac{2r^2}{w_0^2}\right)^{\frac{|l|}{2}} L_p^{|l|}\left(\frac{2r^2}{w_0^2}\right)e^{il\varphi} \tag{8}$$

are the normalized LG modes. Each LG component represents an azimuthal harmonic with orbital number $l$, i.e. $l\hbar$ OAM per photon. $C_{pl}$ is the generalized Fourier coefficient given by

$$C_{pl} = \int_0^{2\pi}\int_0^{\infty} V(r,\varphi,0) u_{pl}^*(r,\varphi,0)\, r\, dr\, d\varphi, \tag{9}$$

that can be explicitly written as

$$C_{pl} = \begin{cases} e^{i(q-l)\pi}\dfrac{\operatorname{Sin}[(q-l)\pi]}{(q-l)\pi}\dfrac{1}{\sqrt{p!(p+|l|)!}}\Gamma\left(\dfrac{|l|}{2}+p\right)\dfrac{|l|}{2} & \text{for } p\ \&\ l \neq 0 \\ e^{iq\pi}\dfrac{\operatorname{Sin}(q\pi)}{q\pi} & \text{for } p\ \&\ l = 0 \end{cases}. \tag{10}$$

Fig. (2) shows the transverse field $z=0$ obtained by diffraction of a Gaussion mode from SPP's with topological charge, respectively, of 0.5,1.5 and 2.5. For $z\neq 0$, the output field can be described by

$$V(r,\varphi,z) = \sum_{p=0}^{\infty} \sum_{l=-\infty}^{\infty} C_{pl} u_{pl}^{LG}(r,\varphi,z), \tag{11}$$

with

$$u_{pl}^{LG}(r,\varphi,z) = \sqrt{\frac{p!}{(p+|l|)!}} \sqrt{\frac{2}{w_z^2 \pi}} \left(\frac{2r^2}{w_z^2}\right)^{\frac{|l|}{2}} L_p^{|l|}\left(\frac{2r^2}{w(z)^2}\right) e^{\left[-\frac{r^2}{w^2(z)} + \frac{ikr^2}{2R(z)}\right]} e^{\left[ikz-(2p+|l|+1)\Phi(z)\right]} e^{il\varphi}. \tag{12}$$

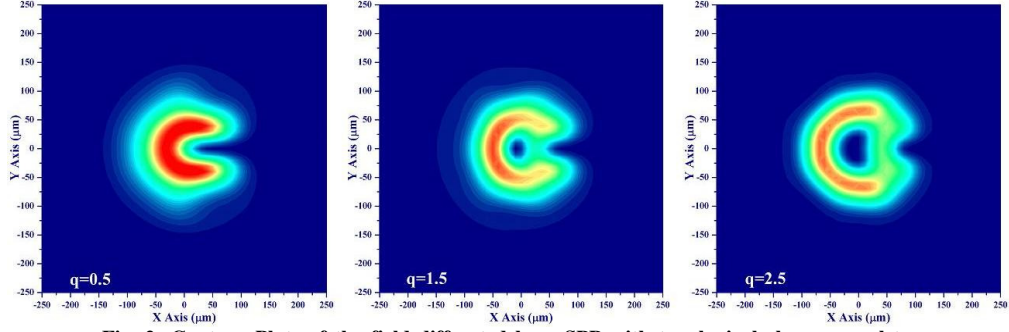

Fig. 2: Contour Plots of the field diffracted by a SPP with topological charge equal to, respectively, 0.5, 1.5 and 2.5.

### 3. Quantum Algebra of Paraxial Beams

A more elegant solution can be derived by considering that a paraxial wave equation has a complete set of solutions in the form of a Gaussian multiplied with a Hermite or Laguerre polynomial, with a scale varying with the transverse plane. These solutions are closely related to the eigenfunctions of the 2-d degenerate quantum Harmonic Oscillator (HO). For the 2-d Isotropic Harmonic Oscillator we can introduce the set of creation and annihilation operators $\hat{a}_j^\dagger = (x_j - d/dx_j)/\sqrt{2}$ and $\hat{a}_j = (x_j + d/dx_j)/\sqrt{2}$ with $j \in \{1,2\}$, which satisfies the usual bosonic commutation rules

$$[\hat{a}_i, \hat{a}_j] = [\hat{a}_i^\dagger, \hat{a}_j^\dagger] = 0 \text{ and } [\hat{a}_i, \hat{a}_j^\dagger] = \delta_{ij}. \tag{13}$$

The number operators are defined as

$$\hat{N}_j = \hat{a}_j^\dagger \hat{a}_j, \tag{14}$$

with

$$\hat{N} = \hat{N}_1 + \hat{N}_2 = \hat{a}_1^\dagger \hat{a}_1 + \hat{a}_2^\dagger \hat{a}_2 \tag{15}$$

being the total number.

We use the basis consisting of tensor products of the one-dimensional eigenvectors, corresponding to the complete set of commuting observables $\{N_1, N_2\}$. The elements of this basis are given by the Hermite–Gaussian modes:

$$|n_1, n_2\rangle = \frac{1}{\sqrt{n_1! n_2!}} \hat{a}_1^{\dagger n_1} \hat{a}_2^{\dagger n_2} |0,0\rangle. \tag{16}$$

The angular momentum L is defined as

$$\hat{L} = i\left(\hat{a}_2^\dagger \hat{a}_1 - \hat{a}_1^\dagger \hat{a}_2\right) \tag{17}$$

and $\{N, L\}$ is another complete set of commuting observables.

By introducing the circular raising operators

$$\hat{a}_\pm^\dagger = \frac{\hat{a}_1^\dagger \pm i\hat{a}_2^\dagger}{\sqrt{2}} \text{ and } \hat{a}_\pm = \frac{\hat{a}_1 \pm i\hat{a}_2}{\sqrt{2}}, \tag{18}$$

and the corresponding number operators

$$\hat{N}_+ = \hat{a}_+^\dagger \hat{a}_+ \text{ and } \hat{N}_- = \hat{a}_-^\dagger \hat{a}_-, \tag{19}$$

we obtain that

$$\hat{N} = \hat{N}_+ + \hat{N}_- = \hat{N}_1 + \hat{N}_2 \tag{20}$$

and

$$\hat{L} = \hat{N}_+ - \hat{N}_-. \tag{21}$$

The elements of the new basis are given by

$$|n_+, n_-\rangle = \frac{1}{\sqrt{n_+! n_-!}} \hat{a}_+^{\dagger n_+} \hat{a}_-^{\dagger n_-} |0,0\rangle. \tag{22}$$

Moreover

$$\left[\hat{L}, \hat{a}_\pm^\dagger\right] = \pm \hat{a}_\pm^\dagger \text{ and } \left[\hat{L}, \hat{a}_\pm\right] = \mp \hat{a}_\pm, \tag{23}$$

So that, when they act on an eigenvector of $\hat{L}$, $\hat{a}_+^\dagger$ and $\hat{a}_-$ increase $L$ by one unit, and $\hat{a}_-^\dagger$ and $\hat{a}_+$ decrease $L$ by one unit.

In complex notation, we can write $z = x + iy$, so that

$$\hat{a}_+^\dagger = \frac{1}{2} z - \frac{\partial}{\partial \bar{z}} = -e^{\frac{z\bar{z}}{2}} \frac{\partial}{\partial \bar{z}} e^{-\frac{z\bar{z}}{2}}, \tag{24}$$

$$\hat{a}_-^\dagger = \frac{1}{2} \bar{z} - \frac{\partial}{\partial z} = -e^{\frac{z\bar{z}}{2}} \frac{\partial}{\partial z} e^{-\frac{z\bar{z}}{2}}, \tag{25}$$

$$\hat{a}_+ = \frac{1}{2} \bar{z} + \frac{\partial}{\partial z}, \tag{26}$$

$$\hat{a}_- = \frac{1}{2} z + \frac{\partial}{\partial \bar{z}}. \tag{27}$$

As usual, the ground state is defined by the solution of the equations

$$\hat{a}_+ |0,0\rangle = 0 \text{ and } \hat{a}_- |0,0\rangle = 0, \tag{28}$$

which give, in complex coordinate representation,

$$\langle z, \bar{z} | 0,0 \rangle = \frac{1}{\sqrt{\pi}} e^{-\frac{z\bar{z}}{2}}. \tag{29}$$

The position states $|z,\bar{z}\rangle$ are orthogonal and form a complete basis in the single-photon Hilbert space. All other eigenvectors are obtained by applying the creation operators to the ground state and correspond to Laguerre-Gaussian modes. Explicitly,

$$\langle z,\bar{z} | n_+, n_-\rangle = \frac{1}{\sqrt{\pi}} \frac{(-1)^{n_+ + n_-}}{\sqrt{n_+! n_-!}} \left(e^{\frac{z\bar{z}}{2}} \frac{\partial}{\partial \bar{z}} e^{-\frac{z\bar{z}}{2}}\right)^{n_+} \left(e^{\frac{z\bar{z}}{2}} \frac{\partial}{\partial z} e^{-\frac{z\bar{z}}{2}}\right)^{n_-} e^{-\frac{z\bar{z}}{2}}, \qquad (30)$$

that, for $n_+ > n_-$, brings to

$$\langle z,\bar{z} | n_+, n_-\rangle = \frac{(-1)^{+n_-}}{\sqrt{\pi}} \sqrt{\frac{n_-!}{n_+!}} z^{n_+ - n_-} e^{-\frac{z\bar{z}}{2}} L_{n_-}^{n_+ - n_-}(z\bar{z}) \qquad (31)$$

and for $n_- \geq n_+$ can be rewritten as

$$\langle z,\bar{z} | n_+, n_-\rangle = \frac{(-1)^{n_+}}{\sqrt{\pi}} \sqrt{\frac{n_+!}{n_-!}} \bar{z}^{n_- - n_+} e^{-\frac{z\bar{z}}{2}} L_{n_+}^{n_- - n_+}(z\bar{z}). \qquad (32)$$

## 4. The Spiral Phase Plate Quantum Operator

A spiral phase plate is a transparent dielectric plate with an edge dislocation that can be freely rotated around the plate axis. Let $z$ be the axis of the plate, $\varphi$ the azimuth angle and without loss of generality, let us suppose that the edge dislocation is at $\varphi = 0$. When a light beam with transverse profile $V(z\bar{z})$ crosses such a SPP, it acquires an azimuthal-dependent phase $e^{iq\varphi}$, with $q \in \mathbb{R}$, i.e., in terms of wave function,

$$\Psi(z\bar{z}) \Rightarrow e^{iq\varphi} \langle z,\bar{z} | \Psi\rangle. \qquad (33)$$

Let us introduce the quantum operator

$$\hat{e}^{i\varphi} = \frac{z}{|z|} = \sqrt{\frac{\hat{a}_+^\dagger + \hat{a}_-}{\hat{a}_+ + \hat{a}_-^\dagger}} = \sqrt{\frac{\hat{Y}}{\hat{Y}^\dagger}}, \qquad (34)$$

with the commutator

$$\left[\hat{Y}, \hat{Y}^\dagger\right] = 0. \qquad (35)$$

The generic operator $\hat{e}^{iq\varphi}$ can be expanded in the form

$$\hat{e}^{iq\varphi} = \sum_{k=-\infty}^{\infty} C_{qk} \hat{e}^{ik\varphi}, \qquad (36)$$

where

$$C_{qk} = \frac{1}{2\pi} \int_0^{2\pi} e^{i(q-k)\varphi} d\varphi = \frac{e^{i(q-k)\pi} Sin[(q-k)\pi]}{(q-k)\pi} \qquad (37)$$

and

$$\hat{e}^{ik\varphi} = \left( \frac{\hat{a}_+^\dagger + \hat{a}_-}{\hat{a}_+ + \hat{a}_-^\dagger} \right)^{\frac{k}{2}} \tag{38}$$

with $k \in \mathbb{N}$.

Following [26], or better, Isaac Newton (1665), the general operator $\hat{e}^{ik\varphi}$ can be formally expanded as

$$\hat{e}^{ik\varphi} = \left( \hat{a}_+^\dagger + \hat{a}_- \right)^{\frac{k}{2}} \left( \hat{a}_+ + \hat{a}_-^\dagger \right)^{-\frac{k}{2}} = \left( \hat{a}_+ + \hat{a}_-^\dagger \right)^{-\frac{k}{2}} \left( \hat{a}_+^\dagger + \hat{a}_- \right)^{\frac{k}{2}}$$

$$= \sum_{m=0}^\infty \sum_{h=0}^\infty \binom{k/2}{m} \binom{-k/2}{h} \hat{a}_+^{\dagger m} \hat{a}_-^{\frac{k}{2}-m} \hat{a}_+^h \hat{a}_-^{\dagger -\frac{k}{2}-h}, \tag{39}$$

where

$$\binom{k/2}{h} = \frac{(k/2)_h}{h!}. \tag{40}$$

$(r)_h$ is the Pochhammer symbol, which is used to represent the falling sequential products (sometimes improperly called *"factorials"*) by using the Gamma function:

$$(r)_h = \frac{\Gamma(1+r)}{\Gamma(1+r-h)}, \tag{41}$$

The Eq.(39) can be considered a novel "two modes" formulation of the Paul's phase operator [16]. By exploiting the Glauber-Sudarshan P-representation, Paul introduced the operator

$$\hat{E}_1 = \hat{e}^{i\varphi} = \frac{1}{\pi} \int d^2\alpha \frac{\alpha}{|\alpha|} |\alpha\rangle\langle\alpha|, \tag{42}$$

and its generalizations

$$\hat{E}_k = \hat{e}^{ik\varphi} = \frac{1}{\pi} \int d^2\alpha \frac{\alpha^k}{|\alpha|^k} |\alpha\rangle\langle\alpha|, \tag{43}$$

$$\hat{E}_{-k} = e^{-ik\varphi} = \frac{1}{\pi} \int d^2\alpha \frac{\alpha^{*k}}{|\alpha|^k} |\alpha\rangle\langle\alpha| \qquad (k=1,2...), \tag{44}$$

in order to include the operators corresponding to powers of the classical quantities $e^{ik\varphi}$ and $e^{-ik\varphi}$, respectively.

Later, Fan Hong-Yi and Min Xiao, starting from the 2-d oscillator in the entangled state representation [21-25], found an expression in which the two-mode phase operator $\hat{e}^{i\varphi}$ manifestly exhibits its phase behavior. Hinted by EPR, the bipartite entangled state $|\xi\rangle$ is introduced as

$$|\xi\rangle = e^{-\frac{|\xi|^2}{2} + \xi a_1^\dagger + \xi^* a_2^\dagger - a_1^\dagger a_2^\dagger} |0,0\rangle, \tag{45}$$

The state $|\xi\rangle$ is the common eigenvector of the center of mass coordinate and the relative momentum in two-mode Fock space, and satisfies the relations

$$\langle\xi'|\xi\rangle = \pi\delta(\xi'-\xi)\delta(\xi'^*-\xi^*) \text{ and } \int\frac{d^2\xi}{\pi}|\xi\rangle\langle\xi| = 1. \tag{46}$$

The operator $\hat{e}^{i\varphi}$ and $\hat{e}^{i\varphi\dagger}$ have the above defined state as eigenvectors. In fact

$$\left(\hat{a}_1 + \hat{a}_2^\dagger\right)|\xi\rangle = \xi|\xi\rangle \text{ and } \left(\hat{a}_2 + \hat{a}_1^\dagger\right)|\xi\rangle = \xi^*|\xi\rangle \tag{47}$$

And, because $\left(\hat{a}_1 + \hat{a}_2^\dagger\right)$ and $\left(\hat{a}_2 + \hat{a}_1^\dagger\right)$ are diagonalized in $|\xi\rangle$, Fan Hong-Yi and Min Xiao obtained

$$\hat{e}^{i\varphi} = \sqrt{\frac{\left(\hat{a}_1 + \hat{a}_2^\dagger\right)}{\left(\hat{a}_2 + \hat{a}_1^\dagger\right)}} = \sqrt{\frac{\xi}{\xi^*}}|\xi\rangle = e^{i\varphi}|\xi\rangle \tag{48}$$

and

$$\hat{e}^{i\varphi\dagger} = \sqrt{\frac{\left(\hat{a}_2 + \hat{a}_1^\dagger\right)}{\left(\hat{a}_1 + \hat{a}_2^\dagger\right)}} = \sqrt{\frac{\xi^*}{\xi}}|\xi\rangle = e^{-i\varphi}|\xi\rangle \tag{49}$$

Moreover, by using the technique of integration within an ordered product (IWOP) of operators, the generic operator $\hat{e}^{ik\varphi}$ can be expressed in the form

$$\hat{e}^{ik\varphi} = \sum_{n=0}^{\infty}\Gamma\left(n+\frac{k}{2}+1\right):\frac{\left(\hat{Y}^\dagger\hat{Y}\right)^n \hat{Y}^\dagger}{n!(n+1)!}e^{-\hat{Y}^\dagger\hat{Y}}: . \tag{50}$$

In the following limit

$$_2\langle 0|\hat{e}^{i\varphi}|0\rangle_2 = \int\frac{d^2\xi}{\pi}e^{i\varphi}\,_2\langle 0|\xi\rangle\langle\xi|0\rangle_2, \tag{51}$$

the operator $\hat{e}^{ik\varphi}$ reduces to the Paul operator.

All these representations of the phase operator circumvent the problem of applying the nonlinear phase operator directly to Fock states by exploiting non-orthogonal states such as Glauber-Sudarshan states or EPR bipartite states.

Usually, Fock states $|n\rangle$ are defined as eigenstates of the photon number operator

$$\hat{N} = \hat{a}^\dagger\hat{a}, \tag{52}$$

and each state, with an integer number of excitations, can be obtained by applying an integer power of the creation operator $\hat{a}^\dagger$ to vacuum, namely

$$|n\rangle = \frac{\hat{a}^{\dagger n}}{\sqrt{n!}}|0\rangle. \tag{53}$$

However, nothing prevents to obtain a generic number state by applying a non-integer power of the operators $\hat{a}^\dagger$ and $\hat{a}$ such as

$$|n\rangle = \frac{\sqrt{\Gamma(1+\beta-\alpha)}}{\Gamma(1+\beta)}\hat{a}^{\alpha}\left(\hat{a}^{\dagger}\right)^{\beta}|0\rangle \tag{54}$$

or

$$|n\rangle = \frac{\Gamma(1-\alpha)}{\sqrt{\Gamma(1+\beta-\alpha)}}\left(\hat{a}^{\dagger}\right)^{\beta}\hat{a}^{\alpha}|0\rangle, \tag{55}$$

if we put $(\beta-\alpha) = n$ and we remember the analytical domain of Gamma function.

Moreover, nothing prevents to apply the operators $\hat{a}^{\alpha}\hat{a}^{\dagger\beta}$ and $\hat{a}^{\dagger\beta}\hat{a}^{\alpha}$ to a generic Fock state by exploiting the analytical properties of Gamma function, namely

$$\hat{a}^{\alpha}\hat{a}^{\dagger\beta}|n\rangle = \frac{\Gamma(1+n+\beta)}{\sqrt{n!\Gamma(1+n+\beta-\alpha)}}|n+\beta-\alpha\rangle, \tag{56}$$

and

$$\hat{a}^{\dagger\beta}\hat{a}^{\alpha}|n\rangle = \frac{\sqrt{n!\Gamma(1+n+\beta-\alpha)}}{\Gamma(1+n-\alpha)}|n+\beta-\alpha\rangle, \tag{57}$$

with the constrain $n+\beta-\alpha \in \mathbb{N}^+$.

We can use these rules to calculate the matrix elements of the operator $\hat{e}^{ik\varphi}$

$$\begin{aligned}
&\langle n'|\langle m'|\hat{e}^{ik\varphi}|m\rangle|n\rangle \\
&= \langle n'|\langle m'|\sum_{i=0}^{\infty}\sum_{h=0}^{\infty}\binom{k/2}{i}\binom{-k/2}{h}\hat{a}_+^{\dagger i}\hat{a}_-^{\frac{k}{2}-i}\hat{a}_+^{h}\hat{a}_-^{\dagger \frac{k}{2}-h}|m\rangle|n\rangle \\
&= \sum_{i=0}^{\infty}\sum_{h=0}^{\infty}\binom{k/2}{i}\binom{-k/2}{h}\frac{\sqrt{m!(m+i-h)!}\,\Gamma\left(1+n-\frac{k}{2}-h\right)}{(m-h)!\sqrt{n!(n+i-h-k)!}}\delta_{n',n+i-h-k}\delta_{m',m+i-h}.
\end{aligned} \tag{58}$$

For $|m\rangle = |0\rangle$ we obtain

$$\begin{aligned}
\langle n'|\langle 0|\hat{e}^{ik\varphi}|0\rangle|n\rangle &= \langle n'|\langle 0|\sum_{i=0}^{\infty}\sum_{h=0}^{\infty}\binom{k/2}{i}\binom{-k/2}{h}\hat{a}_+^{\dagger i}\hat{a}_-^{\frac{k}{2}-i}\hat{a}_+^{h}\hat{a}_-^{\dagger \frac{k}{2}-h}|0\rangle|n\rangle \\
&= \frac{\Gamma\left(1+n-\frac{k}{2}\right)}{\sqrt{n!(n-k)!}}\delta_{n',n-k},
\end{aligned} \tag{59}$$

which proves, in this limit, the equivalence with Paul's operator.

We can also demonstrate that $\hat{e}^{-ik\varphi}$ is the Hermitian conjugate of $\hat{e}^{ik\varphi}$, in fact

$$\begin{aligned}
\langle n'|\langle m'|\hat{e}^{ik\varphi\dagger}|m\rangle|n\rangle &= \langle n'|\langle m'|\sum_{i=0}^{\infty}\sum_{h=0}^{\infty}\binom{k/2}{i}\binom{-k/2}{h}\left(\hat{a}_+^{\dagger i}\hat{a}_-^{\frac{k}{2}-i}\hat{a}_+^{h}\hat{a}_-^{\dagger \frac{k}{2}-h}\right)^{\dagger}|m\rangle|n\rangle \\
&= \langle n'|\langle m'|\sum_{i=0}^{\infty}\sum_{h=0}^{\infty}\binom{-k/2}{h}\binom{k/2}{i}\hat{a}_+^{\dagger h}\hat{a}_-^{-\frac{k}{2}-h}\hat{a}_+^{i}\hat{a}_-^{\dagger \frac{k}{2}-i}|m\rangle|n\rangle \\
&= \langle n'|\langle m'|\hat{e}^{-ik\varphi}|m\rangle|n\rangle.
\end{aligned} \tag{60}$$

To prove the equivalence with the classical result, let us apply $\hat{e}^{iq\varphi} = \sum_{k=-\infty}^{\infty} C_{qk} \hat{e}^{ik\varphi}$ to the vacuum state

$$\sum_{k=-\infty}^{\infty} C_{qk} \left( \frac{\hat{a}_- + \hat{a}_+^\dagger}{\hat{a}_-^\dagger + \hat{a}_+} \right)^{\frac{k}{2}} |0,0\rangle = \sum_{k=-\infty}^{\infty} C_{qk} \left( \hat{a}_- + \hat{a}_+^\dagger \right)^{\frac{k}{2}} \left( \hat{a}_-^\dagger + \hat{a}_+ \right)^{-\frac{k}{2}} |0,0\rangle. \qquad (61)$$

We obtain that

$$\sum_{k=-\infty}^{\infty} C_{qk} \sum_{m=0}^{\infty} \sum_{h=0}^{\infty} \binom{k/2}{m} \binom{-k/2}{h} \hat{a}_+^{\dagger m} \hat{a}_-^{\frac{k}{2}-m} \hat{a}_+^h \hat{a}_-^{\dagger \frac{-k}{2}-h} |0,0\rangle$$

$$\sum_{m=0}^{\infty} \sum_{k=-\infty}^{m} C_{qk} \frac{\Gamma\left(1+\frac{k}{2}\right)\Gamma\left(1-\frac{k}{2}\right)}{\Gamma\left(1+\frac{k}{2}-m\right)} \frac{1}{\sqrt{m!(m-k)!}} |m, m-k\rangle. \qquad (62)$$

Because $\left[ \hat{a}_- + \hat{a}_+^\dagger, \hat{a}_-^\dagger + \hat{a}_+ \right] = 0$, the same result can be also obtained by applying the operator $\hat{e}^{iq\varphi} = \sum_{k=-\infty}^{\infty} C_{qk} \hat{e}^{ik\varphi}$ as

$$\sum_{k=-\infty}^{\infty} C_{qk} \left( \frac{\hat{a}_- + \hat{a}_+^\dagger}{\hat{a}_-^\dagger + \hat{a}_+} \right)^{\frac{k}{2}} |0,0\rangle = \sum_{k=-\infty}^{\infty} C_{qk} \left( \hat{a}_-^\dagger + \hat{a}_+ \right)^{-\frac{k}{2}} \left( \hat{a}_- + \hat{a}_+^\dagger \right)^{\frac{k}{2}} |0,0\rangle$$

$$\sum_{h=0}^{\infty} \sum_{k=-h}^{\infty} C_{qk} \frac{\Gamma\left(1-\frac{k}{2}\right)\Gamma\left(1+\frac{k}{2}\right)}{\Gamma\left(1-\frac{k}{2}-h\right)} \frac{1}{\sqrt{h!(h+k)!}} |h+k, h\rangle, \qquad (63)$$

which is equivalent to Eq. (62) with the substitution $h = m - k$. Thanks to the variable changes $k = n_+ - n_-$ and $m = n_+$, we can write

$$\sum_{k=-\infty}^{\infty} C_{qk} \left( \frac{\hat{a}_- + \hat{a}_+^\dagger}{\hat{a}_-^\dagger + \hat{a}_+} \right)^{\frac{k}{2}} |0,0\rangle = \sum_{n_+=0}^{\infty} \sum_{n_-=0}^{\infty} C_{q(n_+-n_-)} \frac{\Gamma\left(1+\frac{n_+-n_-}{2}\right)\Gamma\left(1-\frac{n_+-n_-}{2}\right)}{\Gamma\left(1-\frac{n_++n_-}{2}\right)\sqrt{n_+!n_-!}} |n_+, n_-\rangle. \qquad (64)$$

By exploiting the properties of Gamma function the above expression can be arranged as

$$\sum_{n_+=0}^{\infty} \sum_{n_-=0}^{\infty} C_{q(n_+-n_-)} \frac{(n_+ - n_-)}{2} \Gamma\left(\frac{n_+ + n_-}{2}\right) \frac{Sin\left[\frac{\pi}{2}(n_+ + n_-)\right]}{Sin\left[\frac{\pi}{2}(n_+ - n_-)\right]} \frac{1}{\sqrt{n_+!}\sqrt{n_-!}} |n_+, n_-\rangle. \qquad (65)$$

As usually, we consider two cases. For $n_+ \geq n_-$, we put $n_- = p$ and $l = n_+ - n_-$, so that

$$\sum_{p=0}^{\infty} \sum_{l=0}^{\infty} \frac{e^{i(q-l)\pi} Sin\left[\pi(q-l)\right]}{\pi(q-l)} \frac{l}{2} \Gamma\left(p + \frac{l}{2}\right) \frac{(-1)^p}{\sqrt{p!p+l!}} \langle z, \bar{z} | p+l, p \rangle$$

$$= \sum_{p=0}^{\infty} \sum_{l=0}^{\infty} \frac{e^{i(q-l)\pi} Sin[\pi(q-l)]}{\pi(q-l)} \frac{l}{2} \Gamma\left(p+\frac{l}{2}\right) \frac{(-1)^{2p}}{\sqrt{p!\,p+l!}} \sqrt{\frac{p!}{\pi(p+l)!}} (z\bar{z})^{\frac{l}{2}} e^{-\frac{z\bar{z}}{2}} L_p^l(z\bar{z}) e^{il\varphi}. \quad (66)$$

For $n_+ < n_-$, we put $n_+ = p$ and $l = n_+ - n_-$, so that

$$\sum_{p=0}^{\infty} \sum_{l=-\infty}^{-1} \frac{e^{i(q-l)\pi} Sin[\pi(q-l)]}{\pi(q-l)} \frac{-l}{2} \Gamma\left(p-\frac{l}{2}\right) \frac{Sin\left[\frac{\pi}{2}(2p-l)\right]}{Sin\left(\frac{\pi}{2}l\right)} \frac{1}{\sqrt{p!\,p-l!}} \langle z, \bar{z} \mid p, p-l \rangle$$

$$= \sum_{p=0}^{\infty} \sum_{l=-\infty}^{-1} \frac{e^{i(q-l)\pi} Sin[\pi(q-l)]}{\pi(q-l)} \frac{|l|}{2} \Gamma\left(p+\frac{|l|}{2}\right) \frac{(-1)^{2p}}{\sqrt{p!\,p+|l|!}} \sqrt{\frac{p!}{\pi p+|l|!}} (z\bar{z})^{\frac{|l|}{2}} e^{-\frac{z\bar{z}}{2}} L_p^{|l|}(z\bar{z}) e^{il\varphi}. \quad (67)$$

By the variables change

$$z\bar{z} = \frac{2r_0^2}{w_0^2}, \quad (68)$$

we obtain the same result of the classical approach.

## 5. Displaced Gaussian Beam

Let us consider a normalized Gaussian beam with amplitude and waist transversal dimension $w_0$, so that the amplitude distribution in the waist attains the form

$$u_{00}^{LG}(x,y,0) = \sqrt{\frac{2}{\pi w_0^2}} e^{-\frac{x^2+y^2}{w_0^2}}. \quad (69)$$

The normalized Gaussian beam displaced at the coordinate values $x_0$ and $y_0$, at $z=0$, can be written as

$$u(x,y,0) = \sqrt{\frac{2}{\pi w_0^2}} e^{-\frac{(x-x_0)^2+(y-y_0)^2}{w_0^2}}. \quad (70)$$

By using the transformations

$$x_0 = r_0 Cos(\varphi_0) \quad \text{and} \quad y_0 = r_0 Sen(\varphi_0), \quad (71)$$

in cylindrical coordinates we obtain

$$u(r,\varphi,0) = \sqrt{\frac{2}{\pi w_0^2}} e^{-\frac{r^2+r_0^2-2rr_0 Cos(\varphi-\varphi_0)}{w_0^2}}. \quad (72)$$

Recalling that

$$e^{zCos(\varphi)} = \sum_{l=-\infty}^{\infty} I_l(z) e^{il\varphi}, \quad (73)$$

we can express a displaced Gaussian Beam as a spectral decomposition in terms of modified Bessel functions $I_l$, each carrying an orbital number $l$:

$$u(r,\varphi,0) = \sqrt{\frac{2}{\pi w_0^2}} e^{-\frac{r^2}{w_0^2}} e^{-\frac{r_0^2}{w_0^2}} \sum_{l=-\infty}^{\infty} I_l\left(\frac{2rr_0}{w_0^2}\right) e^{il(\varphi-\varphi_0)} . \tag{74}$$

The last expression can be expressed in terms of a LG Beams by applying

$$I_\alpha(x) = (-i)^\alpha J_\alpha(ix) \tag{75}$$

and

$$J_\alpha(ix) = (-i)^\alpha \left(\frac{ix}{2}\right)^\alpha \frac{e^{-t}}{\Gamma(\alpha+1)} \sum_{p=0}^{\infty} \frac{L_p^{(\alpha)}\left(\frac{x^2}{4t}\right)(-t)^p}{\binom{p+\alpha}{p} p!} , \tag{76}$$

where $t$ is an arbitrary parameter.

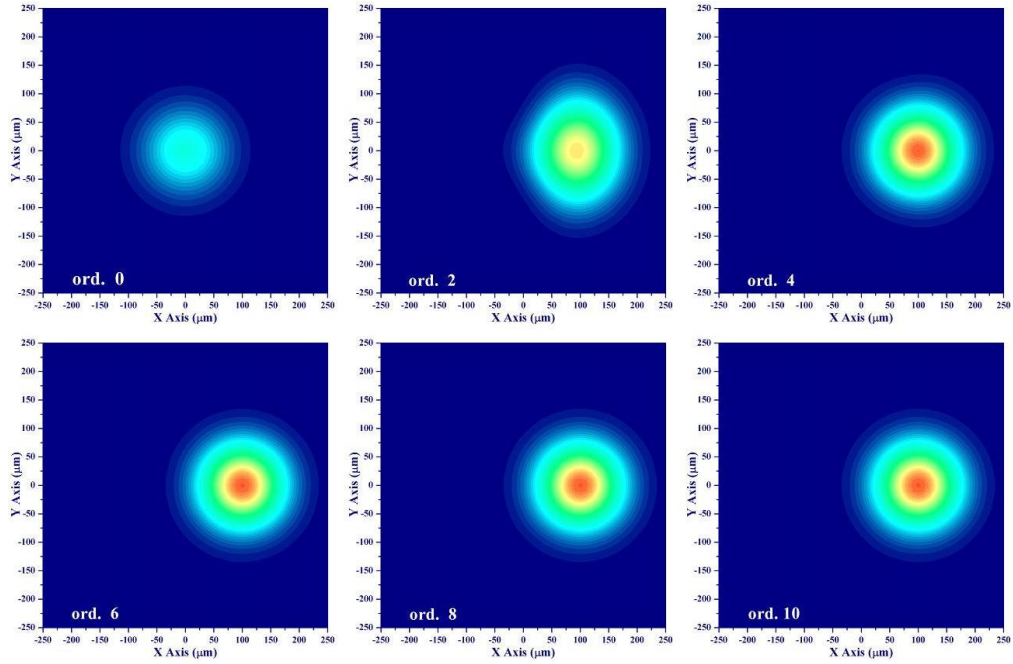

**Fig. 3:** Construction of a Displaced Gaussian Beam starting from the zero order LG component (p=0, l=0). The superposition of the high-order () LG beams definitely produces an off-axis spot. In the particular case, the displacement is equal to the waist, $w_0=100\mu m$.

By choosing

$$x = \frac{2rr_0}{w_0^2} \text{ and } t = \frac{r_0^2}{2w_0^2} , \tag{77}$$

we obtain the spectral decomposition of a Displaced Gaussian Beam in terms of LG Beams:

$$u(r,\varphi,0) = \sum_{p=0}^{\infty} \sum_{l=-\infty}^{\infty} \frac{e^{-\frac{r_0^2}{2w_0^2}}}{\sqrt{p!}} \left(\frac{r_0^2}{2w_0^2}\right)^{p+\frac{|l|}{2}} \frac{(-1)^p}{\sqrt{(p+|l|)!}} u_{pl}^{LG}(r,\varphi-\varphi_0,0). \tag{78}$$

Fig.(3) shows the construction of an off-axis Gaussian Beam by superposing high-order LG Beams.

We can repeat the calculation by using the quantum-mechanics formalism. Let us consider the Displacement Operator $\hat{D}(\alpha_+)\hat{D}(\alpha_-)$ applied to the vacuum state $|0,0\rangle$

$$\hat{D}(\alpha_+)\hat{D}(\alpha_-)|0,0\rangle = e^{\alpha \hat{a}_+^\dagger} e^{\bar{\alpha}\hat{a}_-^\dagger}|0,0\rangle, \tag{79}$$

where

$$\alpha_+ = \bar{\alpha}_- = |\alpha|e^{-i\varphi_0}. \tag{80}$$

We get

$$\langle z,\bar{z}|\hat{D}(\alpha_+)\hat{D}(\alpha_-)|0,0\rangle = \langle z,\bar{z}|\sum_{n_+=0}^{\infty}\sum_{n_-=0}^{\infty} \frac{\alpha^{n_+}\bar{\alpha}^{n_-}}{\sqrt{n_+!n_-!}}|n_+,n_-\rangle. \tag{81}$$

Let us put $p = \min\{n_+, n_-\}$ and $l = n_+ - n_-$. Again, for $n_+ > n_-$ we have that

$$e^{-|\alpha|^2}\sum_{n_-=0}^{\infty}\sum_{n_+=n_-}^{\infty} \frac{\alpha^{n_+}\bar{\alpha}^{n_-}}{\sqrt{n_+!n_-!}} \frac{(-1)^{+n_-}}{\sqrt{\pi}}\sqrt{\frac{n_-!}{n_+!}} z^{n_+-n_-} e^{-\frac{z\bar{z}}{2}} L_{n_-}^{n_+-n_-}(z\bar{z})$$

$$= e^{-|\alpha|^2}\sum_{p=0}^{\infty}\sum_{l=0}^{\infty} \frac{|\alpha|^{2p+l} e^{-il\varphi_0}}{\sqrt{p+l!p!}} \frac{(-1)^p}{\sqrt{\pi}} \sqrt{\frac{p!}{p+l!}} (z\bar{z})^{\frac{l}{2}} e^{-\frac{z\bar{z}}{2}} L_p^l(z\bar{z}) e^{il\varphi}, \tag{82}$$

and for $n_- \geq n_+$ we have that

$$e^{-|\alpha|^2}\sum_{n_+=0}^{\infty}\sum_{n_-=n_+}^{\infty} \frac{\alpha^{n_+}\bar{\alpha}^{n_-}}{\sqrt{n_+!n_-!}} \frac{(-1)^{+n_+}}{\sqrt{\pi}}\sqrt{\frac{n_+!}{n_-!}} \bar{z}^{n_--n_+} e^{-\frac{z\bar{z}}{2}} L_{n_+}^{n_--n_+}(z\bar{z})$$

$$= e^{-|\alpha|^2}\sum_{p=0}^{\infty}\sum_{l=-\infty}^{0} \frac{|\alpha|^{2p+|l|} e^{-il\varphi_0}}{\sqrt{p!p+|l|!}} \frac{(-1)^p}{\sqrt{\pi}} \sqrt{\frac{p!}{p+|l|!}} (z\bar{z})^{\frac{|l|}{2}} e^{-\frac{z\bar{z}}{2}} L_p^{|l|}(z\bar{z}) e^{il\varphi}. \tag{83}$$

By the variables change

$$\alpha = \frac{r_0}{\sqrt{2}w_0} e^{-i\varphi_0}, \quad z\bar{z} = \frac{2r_0^2}{w_0^2} \tag{84}$$

we obtain, again, the classical result described in Eq.(78).

## 6. Displaced Gaussian Beam impinging on a SPP

We can start from the previous result by supposing that an off-axis Gaussian Beam impinges on a SPP. The emerging field in $z = 0$ is given by

$$V(r,\varphi,0) = \sum_{p=0}^{\infty}\sum_{l=-\infty}^{\infty} e^{-\frac{r_0^2}{2w_0^2}} \left(\frac{r_0^2}{2w_0^2}\right)^{p+\frac{|l|}{2}} \frac{(-1)^p e^{-il\varphi_0}}{\sqrt{p!(p+|l|)!}} u_{pl}(r,\varphi,0) e^{im\varphi}. \tag{85}$$

To express the outgoing field in terms of Laguerre-Gaussian Beams, we have to calculate the generalized Fourier coefficients $C_{plhk}$ given by

$$C_{plhk} = \int_0^{2\pi}\int_0^\infty u_{pl}(r,\varphi,0)e^{im\varphi}u_{hk}^*(r,\varphi,0)rdrd\varphi. \tag{86}$$

After straightforward calculations we obtain that

$$C_{plhk} = e^{i(q+l-k)\pi}\sqrt{\frac{p!h!}{(p+|l|)!(h+|k|)!}}\frac{\text{Sin}\left[(q+l-k)\pi\right]}{\pi(q+l-k)}I_{p,h}(l,k), \tag{87}$$

where

$$I_{p,h}(l,k) = \frac{4}{w_0^2}\int_0^\infty e^{-\frac{2r^2}{w_0^2}}\left(\frac{2r^2}{w_0^2}\right)^{\frac{|l|+|k|}{2}}L_p^{(|l|)}\left(\frac{2r^2}{w_0^2}\right)L_h^{(|k|)}\left(\frac{2r^2}{w_0^2}\right)rdr. \tag{88}$$

Thanks to the change of variable

$$x = \frac{2r^2}{w_0^2}, \tag{89}$$

the kernel $I_{p,h}(l,k)$ can be rewritten in the form

$$I_{p,h}(l,k) = \Gamma(\frac{|l|+|k|}{2}+1)\frac{(|l|+1)_p(|k|+1)_h}{p!h!}F_2\left(\frac{|l|+|k|}{2}+1;-p,-h;|l|+1,|k|+1;1,1\right), \tag{90}$$

where $F_2$ is the Appell's function.
By using a standard representation of the Appell's function based on the the Pochhammer symbol $(a)_n$, we can use an alternative expression:

$$I_{p,h}(l,k) = \Gamma(\frac{|l|+|k|}{2}+1)\frac{(|l|+1)_p(|k|+1)_h}{p!h!}\sum_{r=0}^p \frac{(-p)_r\left(\frac{|l|+|k|}{2}+1\right)_r\left(|k|-\frac{|l|+|k|}{2}-r\right)_h}{r!(|l|+1)_r\,(|k|+1)_h}, \tag{89}$$

which is equivalent to

$$I_{p,h}(l,k) = (-1)^{p+h}\Gamma(\frac{|l|+|k|}{2}+1)\sum_{r=0}^{\text{Min}\{p,h\}}\binom{\frac{|k|-|l|}{2}}{p-r}\binom{\frac{|l|-|k|}{2}}{h-r}\binom{\frac{|l|+|k|}{2}+r}{r}. \tag{90}$$

In conclusion, the emerging field in $z=0$ is given by

$$V(r,\varphi,0) = \sum_{p=0}^\infty\sum_{l=-\infty}^\infty e^{-\frac{r_0^2}{2w_0^2}}\left(\frac{r_0^2}{2w_0^2}\right)^{p+\frac{|l|}{2}}\frac{(-1)^p e^{-il\varphi_0}}{\sqrt{p!(p+|l|)!}}\sum_{h=0}^\infty\sum_{k=-\infty}^\infty C_{plhk}u_{hk}(r,\varphi,0)e^{ik\varphi}. \tag{91}$$

In terms of quanto-mechanics, we can repeat the calculation by using the operatorial representation of a SPP. The $LG_0$ mode impinging off-axis on a SPP can be described as

$$\sum_{d=-\infty}^\infty C_{qd}\left(\frac{\hat{a}_-+\hat{a}_+^\dagger}{\hat{a}_-^\dagger+\hat{a}_+}\right)^{\frac{d}{2}}\hat{D}(\alpha_+)\hat{D}(\bar{\alpha}_-)|0,0\rangle$$

$$= \sum_{s=0}^{\infty}\sum_{t=0}^{\infty} \frac{e^{-|\alpha|^2}\alpha^s \bar{\alpha}^t}{\sqrt{s!t!}} \sum_{d=-\infty}^{\infty} C_{qd} \sum_{m=0}^{\infty}\sum_{g=0}^{s} \binom{\frac{d}{2}}{m}\binom{-\frac{d}{2}}{g} \frac{\sqrt{s!(s+m-g)!}\,\Gamma\!\left(1+t-\frac{d}{2}-g\right)}{(s-g)!\sqrt{t!(t+m-g-d)!}} |s+m-g, t+m-g-d\rangle \quad (92)$$

We put $s = p$ and $t - s = l$

$$\sum_{p=0}^{\infty}\sum_{l=-\infty}^{\infty} \frac{e^{-|\alpha|^2}\alpha^p \bar{\alpha}^{p+l}}{(p+l)!} \sum_{d=-\infty}^{\infty} C_{qd} \sum_{m=0}^{\infty}\sum_{g=0}^{p} \binom{\frac{d}{2}}{m}\binom{-\frac{d}{2}}{g} \frac{\sqrt{(p+m-g)!}\,\Gamma\!\left(1+p+l-\frac{d}{2}-g\right)}{(p-g)!\sqrt{(p+m-g+l-d)!}} |p+m-g, p+m-g+l-d\rangle \quad (93)$$

For $l - d = k$ and $p - g = r$ we have that

$$\sum_{p=0}^{\infty}\sum_{l=-\infty}^{\infty} \frac{e^{-|\alpha|^2}\alpha^p \bar{\alpha}^{p+l}}{(p+l)!} \sum_{k=-\infty}^{\infty} C_{q(k-l)} \sum_{m=0}^{\infty}\sum_{r=0}^{p} \binom{\frac{l-k}{2}}{m}\binom{-\frac{l-k}{2}}{p-r} \frac{\Gamma\!\left(1+r+\frac{l+k}{2}\right)\sqrt{(m+r)!}}{r!\sqrt{(m+r+k)!}} |m+r, m+r+k\rangle \quad (94)$$

For $m + r = h$ we obtain

$$\sum_{p=0}^{\infty}\sum_{l=-\infty}^{\infty} \frac{e^{-|\alpha|^2}\alpha^p \bar{\alpha}^{p+l}}{(p+l)!} \sum_{h=0}^{\infty}\sum_{k=-h}^{\infty} C_{q(k-l)} \sum_{r=0}^{\mathrm{Min}\{p,h\}} \Gamma\!\left(1+\frac{l+k}{2}\right) \binom{\frac{l-k}{2}}{h-r}\binom{-\frac{l-k}{2}}{p-r}\binom{\frac{l+k}{2}+r}{r} \sqrt{\frac{h!}{(h+k)!}} |h, h+k\rangle \quad (95)$$

Which, after straightforward calculations, is equivalent to the result described by Eq.(91). Fig.(4) gives a picture of the field generated by a SPP with a topological charge equal to 2.5.

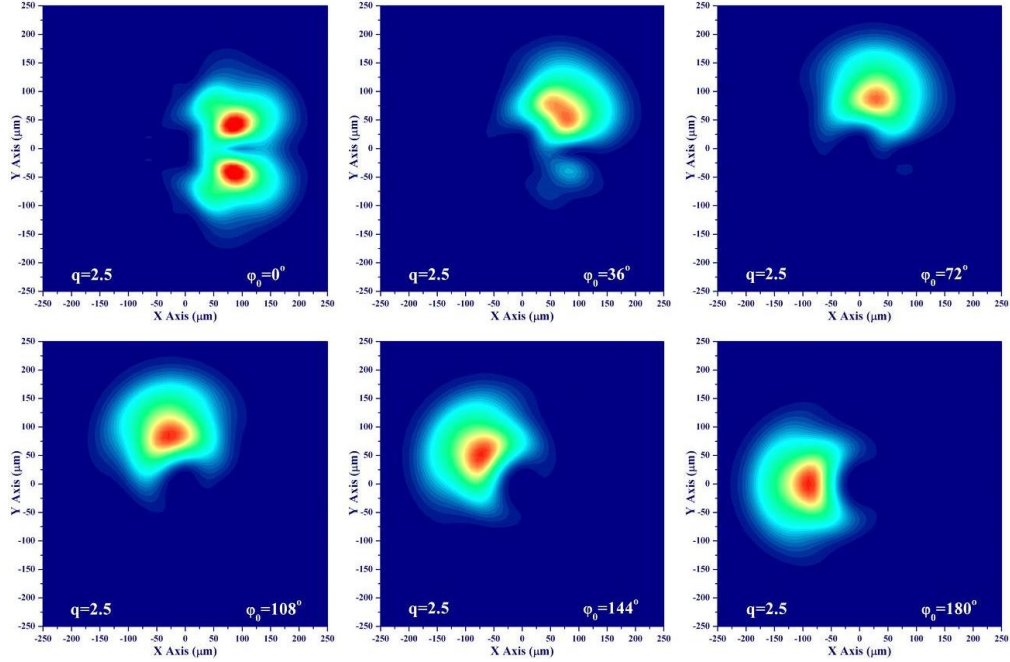

**Fig. 4:** Contour plot of the diffracted field generated by a Displaced Gaussian Beam impinging on a Spiral Phase Plate at different angles in respect to the phase dislocation. The SSP is characterized by a topological charge q=2.5. and, in the particular case, the displacement is equal to the waist, $w_0=100\mu m$.

## 7. Conclusion

Let us summarize the main results of this paper. We identified the quantum operator, representing a Spiral Phase Plate acting on an electromagnetic field, as the two mode phase operator and we introduced a novel representation based on the Newton binomial expansion and on the properties of rational power of lowering and raising operators of electromagnetic field. This method provides to handling nonlinear operators without using auxiliary picture such as P-representation in phase-space. The correctness of this representation is proved by obtaining the same results of the Paul's operator in the single mode limit. Moreover, by exploiting the one-to-one correspondence between the state of the two-dimensional harmonic oscillator and a monochromatic paraxial beam of light, we considered two cases: the action of a Spiral Phase Plate on a $LG_0$ mode (a Gaussian beam corresponding to the vacuum state of the two-dimensional harmonic oscillator) and on a displaced $LG_0$ mode (a displaced vacuum state in quantum picture), obtaining the same results both in classical and quantum picture.

**Acknowledgements**

Thanks to Prof. Giorgio Parisi, Prof. Mario Bertolotti and Prof. Concita Sibilia for critical reading and useful discussions of the manuscript.